\documentclass[preprint2]{emulateapj}
\usepackage{color}

\shorttitle{Water ice in the HD\,100546 disk}
\shortauthors{Honda et al.}

\begin{document}
\title{Water ice at the surface of HD 100546 disk}


\author{M. Honda\altaffilmark{2}, T. Kudo\altaffilmark{3}
S. Takatsuki\altaffilmark{4}, 
A.K. Inoue\altaffilmark{5}, 
T. Nakamoto\altaffilmark{4}, 
M. Fukagawa\altaffilmark{6}, 
M. Tamura\altaffilmark{7,6},
H. Terada\altaffilmark{3} 
and N. Takato\altaffilmark{3} 
}


\altaffiltext{1}{Based on data collected at Subaru Telescope, 
via the time exchange program between Subaru and the Gemini Observatory.
The Subaru Telescope is operated by the National Astronomical Observatory of Japan.}
\altaffiltext{2}{Department of Physics, Kurume University School of Medicine, 
67 Asahi-machi, Kurume, Fukuoka, 830-0011, Japan}
\altaffiltext{3}{Subaru Telescope, National Astronomical Observatory of
Japan, 650 North A'ohoku Place, Hilo, Hawaii 96720, U.S.A.}
\altaffiltext{4}{Department of Earth and Planetary Sciences, Tokyo Institute of Technology, Meguro, Tokyo 152-8551, Japan}
\altaffiltext{5}{College of General Education, Osaka Sangyo University, Daito, Osaka 574-8530, Japan}
\altaffiltext{6}{National Astronomical Observatory of Japan, 2-21-1
Osawa, Mitaka, Tokyo 181-8588, Japan}
\altaffiltext{7}{Department of Astronomy, Graduate School of Science, University of Tokyo, Bunkyo-ku, Tokyo 113-0033, Japan}


\begin{abstract}
\noindent

We made near infrared multicolor imaging observations of a disk around Herbig Be star HD100546 using Gemini/NICI. 
K (2.2\,$\mu$m), H$_2$O ice (3.06\,$\mu$m), and L'(3.8\,$\mu$m) disk images were obtained and we found 
the 3.1\,$\mu$m absorption feature in the scattered light spectrum, likely due to water ice grains at the disk surface. 
We compared the observed depth of the ice absorption feature with the disk model based on \cite{Oka2012} including water ice photodesorption effect by stellar UV photons.
The observed absorption depth can be explained by the both disk models with/without 
photodesorption effect within the measurement accuracy, but slightly favors 
the model with photodesorption effects, implying that the UV photons play 
an important role on the survival/destruction of ice grains at the Herbig Ae/Be 
disk surface.
Further improvement on the accuracy of the observations of the water ice absorption depth is needed to constrain the disk models.
\end{abstract}


\section{Introduction}

Water ice is believed to play many important roles in the planet formation theories.
For example, ice enhances the surface density of solid material in the cold outer part 
of a protoplanetary disk, which promotes the formation of massive cores of gaseous planets \citep[e.g.,][]{Hayashi1985}.
Thus the ice sublimation/condensation front called snowline, 
is considered to be the boundary of the forming regions of the terrestrial and Jovian planets.
Snowline is also suggested as a possible forming 
site of the planetesimals \citep{Brauer2008}. Furthermore, icy planetesimals or comets
may bring water to the Earth \citep[e.g.,][]{Morbidelli2000}. 
Recently, numerous water vapor emission lines have been detected in protoplanetary disks
\citep[e.g.,][]{Carr2008,Salyk2008}, and the position of snow line is inferred from the modeling \citep{Zhang2013}.
On the other hand, observations of water ice distribution in the disk are limited at this moment.
Crystalline H$_2$O ice emission features at 44 and 62\,$\mu$m have
been found for several Herbig Ae/Be stars \citep{Malfait1999,Meeus2001}, but the
limited angular resolution in far-infrared wavelengths hampers us 
to obtain ice distribution. 
While near-infrared (NIR) water ice absorption toward edge-on disks is reported
\citep{Pontoppidan2005,Terada2007}, 
the ice absorption is formed at somewhere through line of sight, thus 
it is still not straightforward to derive its radial distribution in the disk.

\cite{Inoue2008} proposed a new observational way to investigate the radial 
distribution of ice in face-on disks. They showed that ice absorption should 
also be imprinted in the light scattered by icy grains and that multi-wavelength 
imaging in NIR wavebands, including H$_2$O band at 3.1\,$\mu$m, 
is a useful tool to constrain the ice distribution in the disk.
\cite{Honda2009} applied this method to the circumstellar disk around a Herbig
Fe star HD 142527, and showed that the water ice grains present in a disk surface at a
radial distance of 140 AU. On the other hand, \cite{Oka2012} calculated the stability and
distribution of water ice grains in the disk surface considering the photodesorption (photosputtering) process
by UV irradiation. They showed that the water ice grains can be rapidly destroyed at the
disk surface around A/B type stars due to UV photo desorption processes. 
Although \cite{Honda2009} already detected the water ice grains in the disk surface around F-type star
HD142527, it would be interesting to observe the water ice grains in the disk surface around
A/B type stars to check the prediction by \cite{Oka2012}.

In this paper, we showed the observations of water ice grains in the disk surface around
Herbig Be star HD100546, and discuss the presence/stability of water ice grains in the
disk surface. Part of our data was already published in \cite{Currie2014} focusing the planet candidate \citep{Quanz2013} 

\section{Observations and Data Reduction}
Direct imaging observations of the Herbig Be star HD 100546 using K band filter (central
wavelength $\lambda_c$ = 2.20\,$\mu$m, and width $\Delta\lambda$ = 0.33\,$\mu$m), 
H$_2$O ice filter ($\lambda_c$ = 3.06\,$\mu$m, $\Delta\lambda$ = 0.15\,$\mu$m) 
and L' band filter ($\lambda_c$ = 3.78\,$\mu$m, $\Delta\lambda$ = 0.70\,$\mu$m)
were performed using the NICI \citep[Near Infrared Coronagraphic imager ;][]{Chun2008} 
on the Gemini South Telescope on March 31, 2012. 
We fixed the instrument rotator during the observations of both object
and PSF (point-spread function) reference stars to fix the pupil and to obtain 
the stable PSF patterns. A full width at half
maximum (FWHM) of 0.10$''$ was achieved at all the wavelength 
using the instrument AO system. 
The central region close to the central star was saturated, 
however, the outer part (r $>$ 0.22$''$) is
not saturated and can be used for disk observations.

The total exposure times were 1672 s, 3192 s, and 2128 s for K, H$_2$O ice, and L',
respectively. As a PSF reference star, we observed HR 4977 (A0V) just
before/after HD 100546. HD 105116 (K band) and BS4638 (H$_2$O ice and L' band) was
observed as a photometric standard star. The mean 
flux density for BS 4638 at 3.06\,$\mu$m was estimated by scaling the Kurucz's stellar model atmosphere ($T_{eff}$ = 19500 K, $\log g$ = 3.95,
solar metallicity) to match the flux density in K \citep{Soubiran2010}. The calculated flux
density was 5.87 Jy at 3.06\,$\mu$m. Observation parameters are summarized in Table \ref{obssummary}.

\begin{figure}
\epsscale{0.55}
\plotone{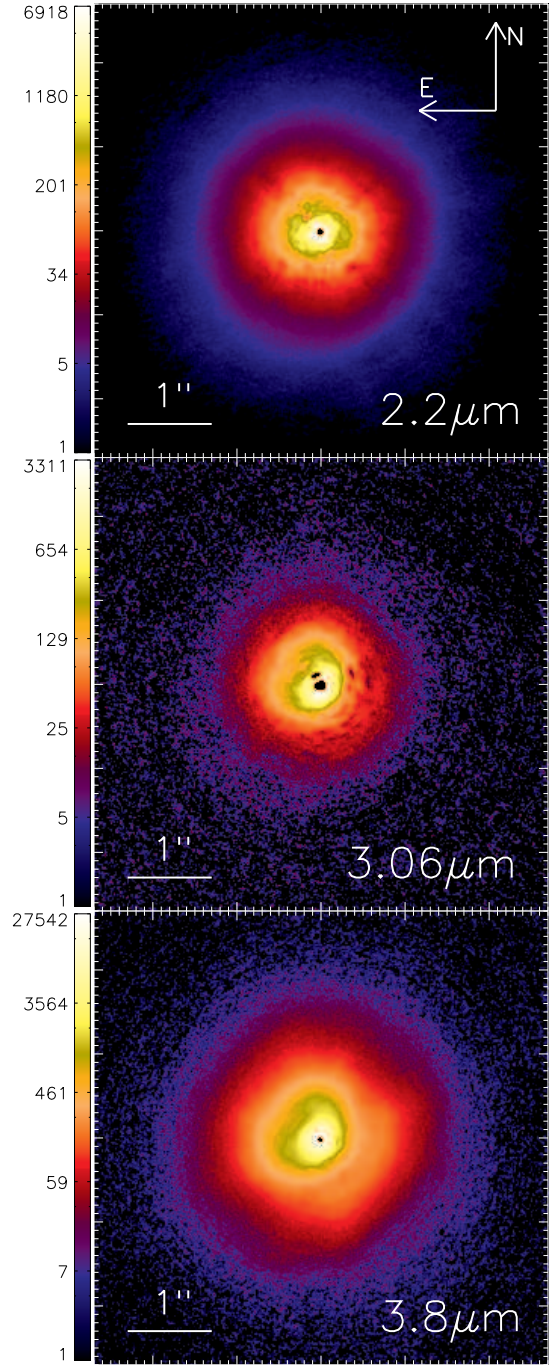}
\caption{
PSF subtracted image of HD 100546 disk at K(top), H$_2$O ice (middle) and L'(bottom). 
Brightness unit is in mJy/arcsec$^2$. North is up and the east is to the left.
}
\label{obsimages}
\end{figure}

The images were first processed using the IRAF packages for dark subtraction,
flat-fielding with sky flats, bad pixel correction, and sky subtraction.
Since the stellar halo was very bright, PSF subtraction was required
to investigate the faint structure near the central star.
The reference PSF was chosen to match the PSF of HD 100546
for each frame with careful visual inspection to determine
whether the circular bright halo of the central point source
was well suppressed after PSF subtraction.
The reference PSF was made by combining the adopted reference star images.
The flux scaling of the reference PSF was performed
so that no region had negative intensity after the subtraction,
in particular, just outer part of central radius (r $>0.25''$).
The reference PSF was shifted to match the central position
and subtracted from each frame of HD 100546.
The each subtracted frame was rotated to match the north direction.
Then, the final PSF subtracted image was made by combining
the rotated object frames.
In order to estimate the systematic uncertainty of the surface brightness
due to the PSF subtraction process, we changed the scaling of the PSF before subtraction, and measured the acceptable range of the scaling factor.
The systematic uncertainty of the surface brightness was
measured to be 20\% depending on the position.
This systematic uncertainty was typically larger than the statistical error
derived from the standard deviation of the best PSF-subtracted object frames.
The final PSF subtracted images of HD100546 disk is shown in Figure \ref{obsimages}.

\section{Results}
In Figure \ref{obsimages}, an extended disk structure is detected in all three bands. Especially in L'
band, a dark lane is seen in the south-west direction from the star, showing the typical
inclined flaring disk morphology. The north-east side is facing to us, while the south-west
scattered light beyond the dark lane can be the scattered light from the other side of the
disk surface. This morphology is consistent with the previous studies \citep{Quanz2011,Mulders2011,Mulders2013,Avenhaus2014}.

\begin{figure}
\epsscale{0.55}
\plotone{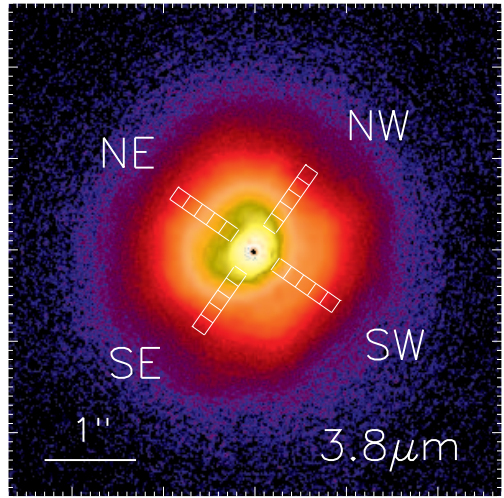}
\caption{
Positions of the spectra extracted region shown in the L' image of HD100546 disk.
A 0.162$''$ square regions were set along the major (SE-NW) and minor (SW-NE) axis of the
disk at the position of 0.360$''$, 0.522$''$, 0.684$''$, 0.846$''$, and 1.008$''$ from the central star.
\label{fig:extract} }
\end{figure}

\begin{figure}
\epsscale{0.54}
\plotone{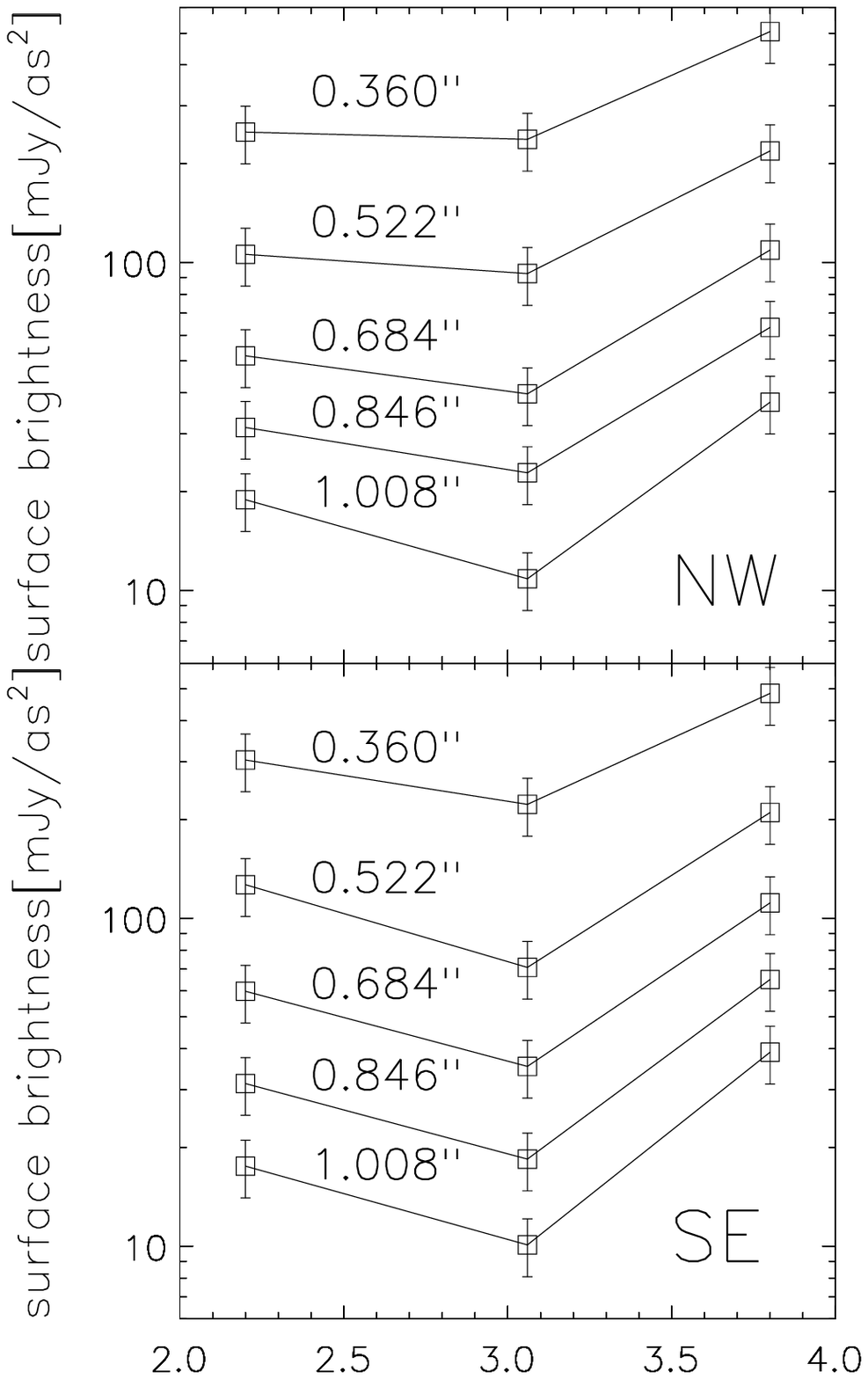}
\plotone{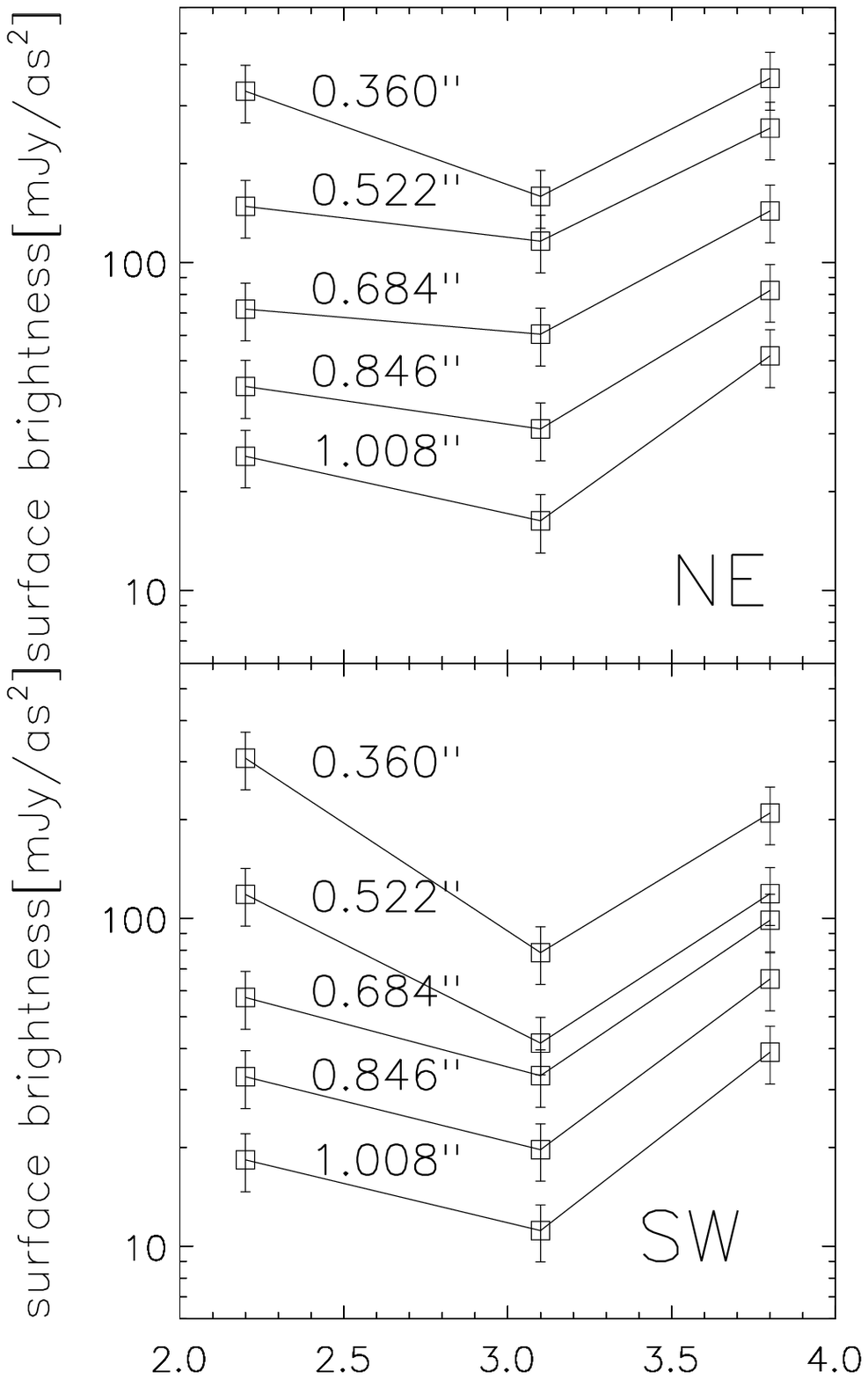}
\caption{\label{fig:spectra}
Extracted spectra along the major (SE-NW) and minor (SW-NE) axis at the
position of 0.360$''$, 0.522$''$, 0.684$''$, 0.846$''$, and 1.008$''$ from the central star shown in Figure \ref{fig:extract}. 
The size of each extracted area is a square region with 0.162$''$ (9 pixels) on the side. In the spectra of
almost all the region, a shallow dip at 3.06\,$\mu$m is seen likely due to water ice absorption.
}
\end{figure}

Using these three color images, we extracted the scattered light spectra of different
region of the protoplanetary disk. Since the position angle (PA) of this disk major axis is
145$^\circ$ \citep{Mulders2011}, we set the 0.162$''$(9 pixel) square region along with major
and minor axis at a distance of 0.360$''$, 0.522$''$, 0.684$''$, 0.846$''$, and 1.008$''$ from the central
star, which is shown in Figure \ref{fig:extract}. The extracted spectra of the each region are shown in
Figure \ref{fig:spectra}. In almost all the regions, relatively shallow 3\,$\mu$m absorption feature is present in their spectra
likely due to water ice grains, indicating that the water ice grains present in the disk surface. 

The shallowness of this ice absorption feature can be due to the loss of ice grains at the 
disk surface. \cite{Oka2012} claimed that the water ice grains can be quickly destroyed 
at the disk surface around A/B type stars due to its strong UV photodesorption. 
To assess their prediction, quantitative comparison with the model prediction is required.

\section{Discussion}

\subsection{3.1\,$\mu$m absorption optical depth $\tau$}

\begin{figure}
\epsscale{0.55}
\plotone{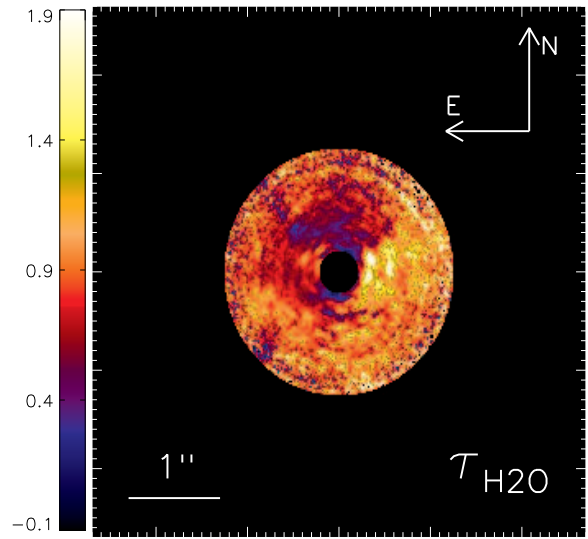}
\caption{\label{fig:H2OTauMap}
$\tau_{H2O}$ map derived from our data. Since the inner region (r$<0.22''$) is
suffered from saturation and the outer region (r$>1.26''$) has low signal 
to noise ratio, these regions are masked.
}
\end{figure}

Since we are interested in the water ice distribution in the disk 
and it is necessary to quantify the absorption feature depth for comparison with the disk model, 
we will use the water ice absorption optical depth $\tau$ following the convention.
Note that the absorption feature in the scattered light spectra is not a pure absorption but rather an albedo effect (see \cite{Inoue2008}), however, 
the optical depth $\tau$ is so often used as an indicator of the depth of the feature, thus we use it for descriptive purposes.
The optical depth is derived from the following formula as usual.

$$\tau = \ln (\frac{I^{cont}_{H_2O}}{I^{obs}_{H_2O}}) $$

where $ I^{obs}_{H_2O}$ is the observed surface brightness at H$_2$O band and 
$ I^{cont}_{H_2O}$ is the estimated continuum surface brightness interpolated 
by power-law using K and L' brightness as shown below.

$$ \log(I^{cont}_{H_2O}) = \frac{\log(I^{obs}_{L'})-\log(I^{obs}_{K})}{\log\lambda_{L'}-\log\lambda_{K}}(\log\lambda_{H_2O}-\log\lambda_K)+\log(I_K^{obs})$$

where $ I^{obs}_{K}$ and $ I^{obs}_{L'}$ are the observed surface brightness at K and L' filters, respectively,
while $ \lambda_{K}$,$ \lambda_{H_2O}$, and $ \lambda_{L'}$ are the central wavelengths of K, H$_2$O and L' filters, respectively.
As the ice absorption depth becomes deeper, the optical depth $\tau$ value becomes larger.
We derived the $\tau$ at each extracted disk regions, and these values ranged from 0 to 2. The $\tau$ map is shown in Fig.\ref{fig:H2OTauMap}. In this figure, the central region ($<$0.22'') is masked due to the saturation problem and outer region (r$>$1.26'') is also masked because of the low signal to noise ratio.
In general, south-western(SW) region shows relatively high $\tau$ value ($\tau\geq 1$), which coincides with the dark-lane seen in L' disk image. This can be qualitatively understood that some scattered light of this region may come from the backside of the disk, 
which suffers more extinction than that from the frontside of the disk.
Other than the SW region, a possible trend could be recognized that the inner region shows lower optical depth ($\tau\leq 1$) which might imply the decrease of the ice grains toward the central star as expected, although the patchy structure in the $\tau$ map hampers us to make a solid conclusion. 

Since the $\tau$ map shows significant asymmetry along the disk minor axis (SW-NE)
and scattered light intensity depends on the scattering geometry, discussion on 
the scattered light spectra along the disk minor axis is more complicated 
than that along the disk major axis (SE-NW). 
Thus we focus on a comparison with the model along the disk major axis.
The radial distributions of $\tau$ along with the disk major axis are summarized in Table\ref{MeasureSummary} and shown in Fig.\ref{fig:H2OTau}. 
As already described, the error is dominated by systematic error, not the statistical one, and is estimated 20\% of the surface brightness at each wavebands.

\subsection{Comparison with the disk model including photodesorption effect}

To discuss whether the observed absorption feature is consistent with
the model predictions,
we derived the expected $\tau$ value based on the disk model
calculations by \cite{Oka2012} who included the effect of
photodesorption of water ice grains by UV photons from the
central star.
Our calculation model consists of two parts:
one is the disk structure calculation (Model Part 1)
in which the density and temperature distributions in the disk,
as well as the snow line, are obtained,
and the other is the radiative transfer calculation (Model Part 2) that simulates observations.

In the Model Part 1, we obtain the location of the snow line,
which is primarily determined by two reasons;
the thermal sublimation and the photodesorption of water ice particles.
In order to evaluate the thermal sublimation of ice particles, we obtain the temperature and
the gas density distributions in the disk as follows.
The temperature in the disk is determined by the energy balance between heating and cooling.
Heating sources for the disk include the radiation from the central star illuminating the surface
of the disk and the viscous heating due to the disk accretion.
Cooling process is the radiative transfer, which finally emits energy from the disk to outer space
by means of radiation.
The surface density distribution is provided as a model parameter, and the gas density distribution
along the vertical direction with respect to the disk is
determined so that the hydrostatic equilibrium is achieved.
The temperature in the disk and the shape of the disk surface affect each other,
because the angle between the direction of the light from the central star and the disk surface
determines the radiative energy received by the disk surface,
and the inclination of the disk surface is a function of the temperature distribution.
Thus, the temperature distribution and the gas density distribution, which is related to the disk
surface, should be calculated consistently.
Using the temperature and the gas density, we can evaluate the vapor pressure
and the saturated vapor pressure of water, which are used to determine the snow line.
On the other hand,
we calculate the radiative transfer of UV radiation from the central star to the disk
and obtain the UV flux exposed to ice particles in the disk,
which gives the photodesorption rate.
Finally,
we can see if ice particles can be present stably with the vapor pressure (contributing to the condensation),
the saturated vapor pressure (contributing to the sublimation),
and the photodesorption rate,
and we obtain the location of the snow line in the disk.

For our calculations, we use physical parameters for HD 100546 given by
\cite{Mulders2011}, which are listed in Table \ref{modelparameter}.
We adopt the mass fractions of silicate and water to the disk gas of 0.0043 and 0.0094,
respectively \cite{MiyakeNakagawa1993}.
We also assume that they form spherical dust particles,
that ice particles contain pure ice and silicate grains are composed of only silicate,
that radii of dust particles range from 0.025 $\mu$m to 2.5 $\mu$m,
and that their size distributions follow the power-law with the index of -3.5.
Absorption and scattering coefficients of dust particles are given by \cite{MiyakeNakagawa1993}.

\begin{figure}
\epsscale{0.55}
\plotone{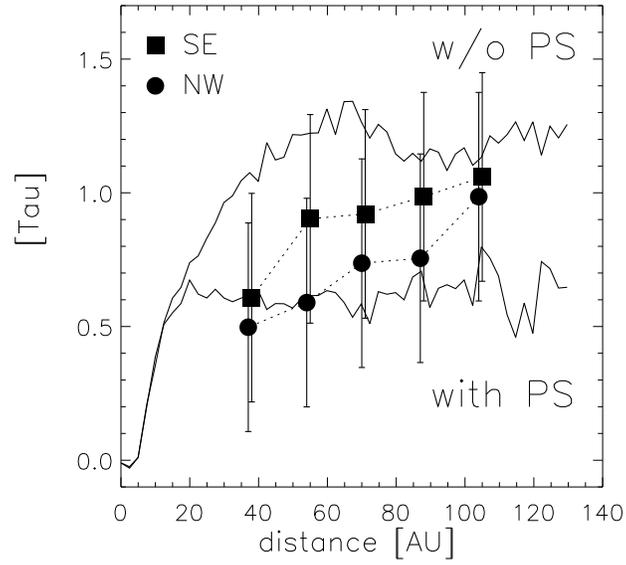}
\caption{\label{fig:H2OTau}
Solid lines are radial profiles of the optical depth $\tau$ of disk models based on \cite{Oka2012} with or without photodesorption (photosputtering; PS) 
along the disk major axis. Over-plotted points are the measured $\tau$ along the disk major axis (SE, NW direction). SE data points are slightly shifted (+1 AU) to avoid the overlapping with NW error-bars.
Both disk models are consistent with the measurements, however, 
the model with photodesorption effect ('with PS') might show slightly better match 
with the observations.
}
\end{figure}

To make a comparison with observation (Model Part 2), we carry out Monte Carlo
radiative transfer calculations.
Deviations of obtained results seen in Fig. \ref{fig:H2OTau} are caused by statistical errors intrinsically
related to Monte Carlo simulations.
In the calculations, we take into account the anisotropic scattering and polarization
produced by dust particles.
The scattering matrices of silicate and ice particles are obtained using the BHmie code
by \citep{BohrenHuffman1983} with the complex refractive indices by \cite{MiyakeNakagawa1993}.

The radial profiles of the $\tau$ of disk models with or without photodesorption
along with the disk major axis are also shown in Fig.\ref{fig:H2OTau}.
A model without photodesorption effect shows deeper water ice absorption feature 
(larger $\tau$) than the observations, 
while models with photodesorption effect show relatively shallow absorption (smaller $\tau$).
This is due to the destruction of water ice grains at the disk surface 
via photodesorption by strong irradiation of UV photons from the central star \citep{Oka2012}.
Although the observed $\tau$ values match with both disk models with/without 
photodesorption effect, the model with photo desorption effect seems 
slightly better match with the observations at least for NW region.

\subsection{Future prospects}

It would be interesting to note that the water vapor is reported to be 
depleted in the disk atmosphere of Herbig AeBe stars by the observations of 
water vapor lines \citep{Fedele2011}.
A plausible explanation is due to photodissociation of water by UV photons 
in the disk atmosphere.
Although photodissociation effect is not included in the model of \cite{Oka2012}, 
they discussed that the photodesorption process is much important 
for the water ice stability, while photodissociation effect is crucial 
for water vapor destruction in the disk surface.
In any cases, the UV photons seem to play important role on the survival 
of both water ice and gas in the disk surface around Herbig Ae/Be stars.

It is apparent that our data do not have a high signal-to-noise ratio enough 
to distinguish the models with/without photodesorption process. 
This is because the systematic error dominates over the total error.
Further observations with better photometric accuracy are 
strongly desired. Since these observations shown here employed 
techniques similar to so-called 'lucky imaging' technique, improvement of 
systematic error is principally limited. However, when we make use of the polarimetric 
differential imaging (PDI) and/or spectroscopy, the systematic error can be significantly
reduced and it will change the situation dramatically.
Thus the L-band PDI and/or spectroscopy is promising for the advance of this observations.
Furthermore, 
other effects on the depth of water ice absorption, 
such as grain size, grain shape, grain structure, ice/rock ratio (abundance),
 dust settling, turbulent mixing, and so on, 
should be investigated in the future theoretical studies 
to comprehensively understand the water ice distribution in the protoplanetary disks.

\acknowledgments

We are grateful to all of the staff members of the Subaru Telescope, 
Gemini Telescope, and Optical Coatings Japan for the production of narrow band filters.
We also thank anonymous referee for their useful comments.
We appreciate Dr. Tom Hayward on his kind support during the observations and
installation of our H$_2$O ice filter to NICI. MH was supported by 
JSPS/MEXT KAKENHI (Grant-in-Aid for Young Scientists B: 21740141, 
Grant-in-Aid for Scientific Research on Innovative Areas: 26108512).
MT is partly supported by the JSPS fund (No. 22000005).




\begin{deluxetable}{ccccc}
\tabletypesize{\scriptsize}
\tablecaption{Observations Summary \label{obssummary}}
\tablewidth{0pt}
\tablehead{
\colhead{object} & \colhead{filter} & \colhead{Integ. time [s]} & \colhead{comment} }
\startdata
HD100546         &  K               & 1672     &  \\
HR4977           &  K               & 456      & PSF reference \\
HD105116         &  K               & 152      & photometric reference  \\
HD100546         &  H$_2$O ice      & 3192     &  \\
HR4977           &  H$_2$O ice      & 2280     & PSF reference \\
BS4638           &  H$_2$O ice      & 152      & photometric reference  \\
HD100546         &  L'              & 2128     &  \\
HR4977           &  L'              & 532      & PSF reference \\
BS4638           &  L'              & 152      & photometric reference
\enddata
\end{deluxetable}

\begin{deluxetable}{rrrrr}
\tabletypesize{\scriptsize}
\tablecaption{Measured Surface Brightness \& Optical Depth $\tau$ along disk major axis\label{MeasureSummary}}
\tablewidth{0pt}\tablehead{
\colhead{distance[AU]} & \colhead{$I^{obs}_{K}$\tablenotemark{*}} & \colhead{$I^{obs}_{H_2O}$\tablenotemark{*}} & \colhead{$I^{obs}_{L'}$\tablenotemark{*}} & \colhead{$\tau$} }
\startdata
\multicolumn{5}{l}{SE} \\
104 & 17.5$\pm$3.5 & 10.0$\pm$2.0 & 39.0$\pm$7.8 & 1.06$\pm$0.39 \\
87 & 31.3$\pm$6.3 & 18.4$\pm$3.7 & 65$\pm$13 & 0.99$\pm$0.39 \\
70 & 60$\pm$12 & 35.3$\pm$7.1 & 111$\pm$22 & 0.92$\pm$0.39 \\
54 & 126$\pm$25 & 71$\pm$14 & 210$\pm$42 & 0.90$\pm$0.39 \\
37 & 304$\pm$61 & 223$\pm$45 & 485$\pm$97 & 0.61$\pm$0.39 \\
\multicolumn{5}{l}{NW} \\
104 & 18.9$\pm$3.8 & 10.8$\pm$2.2 & 37.5$\pm$7.5 & 0.99$\pm$0.39 \\
87 & 31.4$\pm$6.3 & 22.8$\pm$4.6 & 63$\pm$13 & 0.76$\pm$ 0.39 \\
70 & 52$\pm$10 & 39.7$\pm$7.9 & 109$\pm$22 & 0.74$\pm$ 0.39 \\
54 & 106$\pm$21 & 92$\pm$18 & 219$\pm$44 & 0.59$\pm$ 0.39 \\
37 & 250$\pm$50 & 237$\pm$47 & 506$\pm$100 & 0.50$\pm$ 0.39
\enddata
\tablenotetext{*}{in mJy/arcsecond$^2$}
\end{deluxetable}

\begin{deluxetable}{lc}
\tablecaption{Parameters used in the model \label{modelparameter}
}
\tablehead{
\colhead{parameter} & \colhead{values}
}
\startdata
stellar effective temperature $T_*$ &  10500 K       \\
stellar mass $M_*$                  &  2.4 $M_\odot$ \\
stellar luminosity $L_*$            &  36 $L_\odot$  \\
stellar radius $R_*$                &  1.8 $R_\odot$ \\
distance d                          & 103 pc         \\ 
\hline
disk inner radius $r_{in}$          & 0.5 AU         \\
disk outer radius $r_{out}$         & 500. AU        \\
disk inclination                    & 45$^\circ$     \\
surface density at 1AU              & 45 g/cm$^2$    \\ 
surface density power-law index q   & 1              
\enddata
\end{deluxetable}

\end{document}